\documentclass{Interspeech2024}

\usepackage{multirow}
\usepackage[subrefformat=parens]{subcaption}
\newcommand{\mmedit}[1]{\textcolor{black}{#1}}

\interspeechcameraready 

\title{\mmedit{An Attribute} Interpolation Method in Speech Synthesis by Model Merging}

\name{Masato}{Murata}
\name{Koichi}{Miyazaki}
\name{Tomoki}{Koriyama}

\address{
  CyberAgent, Inc., Japan}
\email{murata\_masato@cyberagent.co.jp, miyazaki\_koichi\_xa@cyberagent.co.jp, koriyama\_tomoki@cyberagent.co.jp}


\begin{document}

\maketitle

\begin{abstract}    
With the development of speech synthesis, recent research has focused on challenging tasks, such as speaker generation and emotion intensity control. Attribute interpolation is a common approach to these tasks. However, most previous methods for attribute interpolation require specific modules or training methods. We propose an attribute interpolation method in speech synthesis by model merging. Model merging is a method that creates new parameters by only averaging the parameters of base models. The merged model can generate an output with an intermediate feature of the base models. This method is easily applicable without specific modules or training methods, as it uses only existing trained base models. We merged two text-to-speech models to achieve attribute interpolation and evaluated its performance on speaker generation and emotion intensity control tasks. As a result, our proposed method achieved smooth attribute interpolation while keeping the linguistic content in both tasks.

\end{abstract}

\section{Introduction}

With the development of deep learning, text-to-speech (TTS) systems have achieved human-level naturalness.
Recent research has focused on more challenging tasks for generating variety types of voices, such as speaker generation~\cite{tacospawn, mid_attribute}, which involves generating speech with nonexistent speaker attributes, and emotion intensity control~\cite{lei2022msemotts}, which aims to control the intensity of emotions in a synthesized speech.
These technologies can be applied to the character voice creation in video or game contents\mmedit{, conversational assistants, and audiobook narration.}
Various approaches have been proposed for these tasks, including global style token (GST)-based method~\cite{stanton2018predicting, wu2019end}, speaker/emotion embedding-based method~\cite{tacospawn, mid_attribute}, and Variational Autoencoder (VAE)-based methods~\cite{vaeloop}.
To generate a nonexistent speaker or control the emotion intensity, another common approach is to use an \mmedit{attribute} interpolation method, which \mmedit{perform interpolation} between two base attributes, such as speaker or emotion style. 
However, most conventional attribute interpolation methods require specific training methods or modules, such as introducing speaker/emotion classifier loss~\cite{lei2022msemotts} or pre-trained speaker/emotion encoders~\cite{tacospawn, mid_attribute, emotionlabel}.

\mmedit{Therefore, we focus on a model merging method as} 
an \mmedit{attribute} interpolation 
approach that does not require any specific training methods or modules.
Model merging is a method that creates new model parameters
by simply averaging the parameters of multiple base models \mmedit{trained with different attributes}~\cite{model_soup}.
In most cases, the base models have the same model architecture but are trained with different configurations or datasets.
Recent studies~\cite{gan_cocktail} have shown that a merged model between two generative models can generate an output with an intermediate feature of base models.
This method is easily applicable
because it uses only the \mmedit{existing} trained base models.
Hence, it also does not require preparing the target dataset and computing resources for
training.
The recent trend of openly sharing weight parameters enhances the utility of the model merging method.
With the development of large machine learning models such as the large language model~\cite{gpt} and diffusion-based text-to-image model~\cite{ddpm, latentdiffusion}, many fine-tuned weights are distributed via some platforms such as huggingface\footnote{\url{https://huggingface.co/}} without their datasets.
Considering this trend, the model merging method can be more commonly used because it only requires distributed trained weights to apply with no need to target training data or computing resources.
\mmedit{Therefore, now users can create their original models by easily applying model merging to the existing distributed weights on such platforms.}

\begin{figure}[t]
\vspace{8pt} %
 \centering
  \begin{minipage}[b]{1\linewidth}
    \centering
    \includegraphics[width=\hsize]{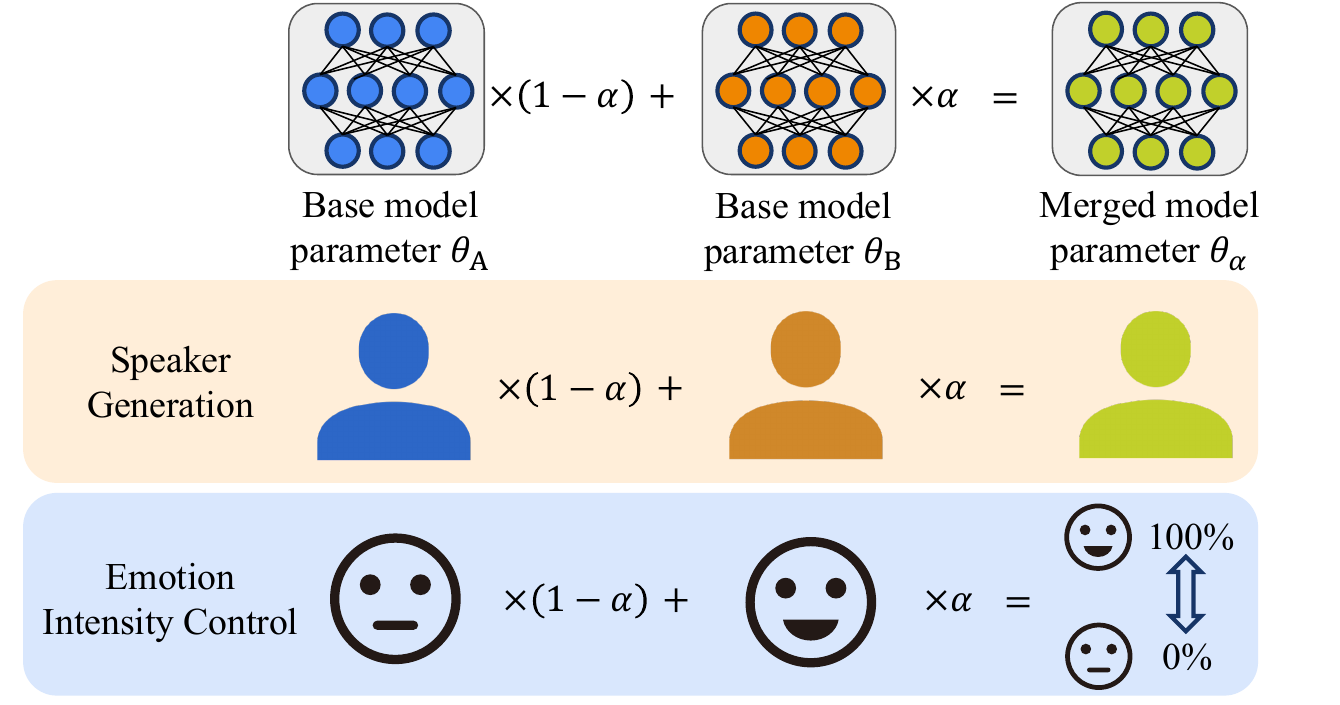}
  \end{minipage}
      \vspace{-15pt} %
  \caption{Overview of the attribute interpolation method by model merging.}
  \label{fig:overview}
  \vspace{-10pt} %
\end{figure}

In this research, to manipulate speech-related attributes, we propose an \mmedit{attribute} interpolation method for TTS models.
Our method merges two TTS models as base models while adjusting the merging coefficient.
Inspired by the capability of the merged model to generate the output with the intermediate feature, we \mmedit{consider} that the merged TTS model can be used for speaker generation and emotion intensity control tasks.
This is because the intermediate acoustic feature can represent the intermediate speaker/emotion between the two base models.
To evaluate the performance of our proposed method, we conduct experiments on two tasks: speaker generation and emotion intensity control.
The experimental results show that our proposed method achieves smooth \mmedit{attribute} interpolation 
while maintaining the linguistic content.
Audio samples are available on our demo page\footnote{\url{https://muramasa2.github.io/model-merge-tts/}}.

Our contributions of this study are as follows:
\begin{itemize}
\item We propose an \mmedit{attribute} interpolation method in speech synthesis \mmedit{by model merging}.
\item In the two experiments: speaker generation and emotion intensity control, we demonstrate the capability of our proposed method across various metrics in the speech domain.
\end{itemize}

\section{Related work}
\subsection{Speaker Generation}

Stanton et al.~\cite{tacospawn} defined a task to synthesize speech of a nonexistent speaker as ``speaker generation".
Speaker generation can be used for audiobook readers, speech assistants, and character voice creation in video or game content.
For the task, several deep neural networks (DNN)-based approaches have been proposed in previous studies.
There are two major approaches, a sampling-based method that samples latent speaker embeddings from learned speaker distributions~\cite{tacospawn, mid_attribute}, and an attribute interpolation-based method that performs speaker interpolation between two reference speakers.

As the attribute interpolation-based methods, Saito et al.~\cite{saito2021perceptual} and Kerle et al.~\cite{xvector-interpolation} used linear interpolation of two base speaker embeddings obtained from a speaker encoder trained on speaker verification task.
Flowtron~\cite{valle2021flowtron} and VAE-Loop~\cite{vaeloop} used latent variable interpolation to generate new speaker features.

In this study, we use a naive attribute interpolation-based method as a baseline which conditions the model on the linear interpolation of base speaker embeddings to compare with the proposed method.

\subsection{Emotion Intensity Control}

Emotion intensity control is a task that aims to control the intensity of emotion in synthesized speech from 0 to 100 \%.

Wang et al. proposed GST, which learns style representation~\cite{gst}.
They controlled the intensity of a specific emotion by scaling the extracted GST embeddings.
To control the emotion intensity, there are other several approaches including learning an emotion embedding space by using an external voice emotion recognition model~\cite{emotionlabel}, and using soft emotion labels obtained by a trained ranking function~\cite{lei2022msemotts}.

\subsection{Model Merging Method}
\label{sec:model_merging_method}

Model merging is a method that creates new parameters by simply averaging the parameters of multiple base models trained with different hyperparameter configurations or datasets.
Let Train$\left(\theta_0, h, d\right)$ denote training from the initial parameter $\theta_0$ with hyperparameter configuration $h$ and target dataset $d$.
The merged model parameters are defined as follows:
\begin{align}
\label{equation:model_merge}
\mathbf{\theta}_m := \frac{1}{k} \sum^k_i{\mathbf{\theta}_i}, &&
\mathbf{\theta}_i = \text{Train}(\mathbf{\theta}_0, h_i, d_i)
\end{align}
where $\mathbf{\theta}_m$ is the parameters of the merged model, $\mathbf{\theta}_0$ is the initial parameters, and $k$ is the number of base models.
The notations $h_i$ and $d_i$ denote the $i$-th hyperparameter configuration and target dataset out of $k$.

Wortsman et al.~\cite{model_soup} proposed the model merging method in discriminative tasks.
They showed that the merged model achieved better accuracy and robustness than each base model, regardless of the architectures of the image and text classification tasks. 
Even in cross-dataset settings, \mmedit{they merged multiple base models which are trained on several datasets such as CIFAR-10~\cite{cifar10} and Food-101~\cite{food100}, and demonstrated that} the merged model outperformed each base model performance in both in-domain and zero-shot scenarios.

Avrahami et al.~\cite{gan_cocktail} applied the model merging method to the image generation task.
\mmedit{They merged two base models that are trained on the datasets belonging to different classes from LSUN~\cite{yu15lsun} dataset, such as cat and dog.}
They demonstrated that the merged model can generate an output including an intermediate \mmedit{semantic} feature between the features of base models.
Thus, this method is easily applicable without any specific training method or modules for \mmedit{attribute} interpolation, as it only uses the existing trained base model. Hence, it also does not require preparing the target dataset and computing resources for training.

\section{Attribute Interpolation Method\\by Model Merging}

Our proposed \mmedit{attribute} interpolation method is based on model merging shown in Figure~\ref{fig:overview}.
Inspired by the characteristics of the merged model being able to generate the output with the intermediate feature between two base models, we consider that the merged TTS model
could be used for speaker generation and emotion intensity control tasks because the intermediate acoustic feature can represent the intermediate speaker/emotion feature between two base models.
In this study, we introduce a simple weighted average model merging between two base models.
\mmedit{This method enables us to create new speakers or control the emotion intensity by merging two base model attributes while controlling the merging coefficient.
Since it does not require any additional training, there is no need to access the target dataset and computing resources because we only use shared trained models as base models.}
The \mmedit{attribute} interpolation method by model merging between the base TTS models A and B is defined as follows:
\begin{align}
\mathbf{\theta}_\alpha := (1-\alpha) \mathbf{\theta}_A + \alpha \mathbf{\theta}_B, &&
0 \leq \alpha \leq 1
\end{align}
where $\alpha$ is the merging coefficient, $\mathbf{\theta}_A$ and $\mathbf{\theta}_B$ are the existing trained weight parameters of each base TTS model.

\section{Experiments}
In this paper, we verified the performance of the proposed method on two tasks: speaker generation and emotion intensity control.
For speaker generation, we used two different single-speaker TTS models as base models A and B.
For emotion intensity control, we used the Neutral style TTS model as base model A and its emotion style TTS model as base model B.
\subsection{Experimental settings}
\label{sec:experimental_settings}
We used two different datasets for each experiment but used the same model architecture for both experiments.

\subsubsection{Datasets}
\label{section:datatset}

In this research, we used two datasets for each experiment: the VCTK~\cite{vctk} dataset for the speaker generation experiment and the ESD~\cite{zhou2021emotional,zhou2021seen} dataset for the emotion intensity control experiment.

\textbf{VCTK}: VCTK~\cite{vctk} dataset contains approximately 44 hours of English speech from 109 individual speakers.
We designed training/validation/test splits of approximately 90\%/5\%/5\% of the original dataset.
For the speaker generation experiment, we first created a multi-speaker TTS model conditioned with speaker IDs as a pre-trained model by training with the VCTK training set.
The detailed training process is referred to the training recipe of ESPnet\footnote{\url{https://github.com/espnet/espnet/blob/master/egs2/vctk/tts1/run.sh}}.
Subsequently, we chose 10 speakers from the VCTK dataset \mmedit{as target speaker for each base model}, which consists of 5 female speakers (speaker IDs: p225, p228, p229, p230, and p231) and 5 male speakers (p226, p227, p232, p237, and p241).
We fine-tuned the pre-trained model with these speakers' data to create each single-speaker TTS model.

\textbf{ESD}: ESD~\cite{zhou2021emotional,zhou2021seen} dataset contains approximately 29 hours of English and 30 hours of Chinese speech uttered by 10 native English and Chinese speakers with 5 emotion styles (Neutral, Angry, Happy, Sad, and Surprise).
For emotion intensity control experiment, we first used the English speaker subset (speaker IDs: 0011--0020) to create a multi-speaker TTS model conditioned with speaker IDs and emotion style IDs as a pre-trained model.
Following that, we chose 5 speakers (0014, 0015, 0018, 0019, and 0020) with 5 emotion styles to fine-tune the pre-trained model to create each speaker's emotion style TTS model.

In both experiments, we used these fine-tuned models as existing trained base models.

\subsubsection{Model Architecture}
We emploied the implementation of Conformer-FastSpeech2 (CFS2)~\cite{cfs2} from ESPnet~\cite{espnet} with the configurations\footnote{\url{https://github.com/espnet/espnet/blob/master/egs2/vctk/tts1/conf/tuning/train_xvector_conformer_fastspeech2.yaml}}.
In TTS applications, single-speaker models are commonly created by fine-tuning from a multi-speaker pre-trained model.
Therefore, in this study, we created each base model fine-tuned from a multi-speaker pre-trained model.
As a pre-trained model, we prepared a speaker/emotion style ID-conditioned multi-speaker TTS model.

Following the prior research~\cite{saito2021perceptual, xvector-interpolation}, we also used the speaker embedding conditioning model with a speaker encoder module as a baseline.
\mmedit{As a speaker embedding, we used x-vector~\cite{xvector} embedding obtained from pre-trained speaker encoder of speechbrain\footnote{\url{https://speechbrain.github.io/}} which is trained on VoxCeleb 1~\cite{voxceleb1} and VoxCeleb2~\cite{voxceleb2}.}
For waveform generation, we used the pre-trained HiFi-GAN~\cite{hifigan} vocoder (vctk\_hifigan.v1) on ParallelWaveGAN repository\footnote{\url{https://github.com/kan-bayashi/ParallelWaveGAN}} trained with the VCTK dataset.

\subsection{Speaker generation}
\label{sec:speaker_interpolation}

Our proposed method performs speaker generation by an attribute interpolation-based approach, called speaker interpolation.
To investigate the performance of the proposed method in the speaker generation task, we merged base models on several speaker combination settings, and evaluated the synthesized speech of each merged model.
Specifically, we first prepared three types of speaker-gender combinations using speaker subsets of the VCTK corpus: Female-Female, Male-Male, and Male-Female.
The female speaker subset consists of the five female speakers, and the male speaker subset also consists of the five speakers \mmedit{described in Section~\ref{section:datatset}.}
Then, we created each single-speaker base model by fine-tuning from the same pre-trained model.
By merging these base models, we generated the speech of nonexistent speakers.
We evaluated the generated speech of the merged models from two perspectives: speech quality and speaker interpolation smoothness.

\subsubsection{Speech quality evaluation}
\label{sec:mos_wer_test}

\mmedit{Since it is not guaranteed that the output of merged models can keep the naturalness and linguistic content of base models,} we evaluated the synthetic speeches generated by merged models based on their naturalness and capability to maintain linguistic content (intelligibility).
In this experiment, we used the mean opinion score (MOS) for naturalness evaluation and the word error rate (WER) for intelligibility evaluation.
We evaluated the synthesized speech output of the merged model ($\alpha$ = 0.5) and that of each base model.
\mmedit{We also evaluated the baseline model conditioned with the average speaker embeddings of two base speakers.
These models are denoted} as ``model merge" and ``spk emb" respectively.

For the MOS test, participants rated the naturalness of the given speech samples on a 5-point scale.
For each gender combination, we randomly selected two base speakers from each gender speaker subset and generated speech samples of interpolated speakers and base speakers.
We asked 40 participants to evaluate five different generated speeches (including interpolated speaker's speech and base speaker's speech).
The sentence of each sample was randomly selected from 10 test sentences in the VCTK test set.
Participants were recruited from the crowdsourcing site Amazon Mechanical Turk (AMT)\footnote{\url{https://www.mturk.com/}}.
For the intelligibility evaluation, we calculated the WER score by using a pre-trained Conformer model~\cite{conformer} on espnet model zoo\footnote{\url{https://github.com/espnet/espnet_model_zoo}}.
\begin{table}[t!]
    \begin{center}
    \caption{MOS score with 95\% confidence interval (CI) and WER score with several gender combination settings.}
    \scalebox{0.68}{
        \begin{tabular}{ c|cccc } 
            \toprule
            \multicolumn{5}{c}{MOS} \\
            \toprule
             \multirow{2}{*}{Speaker combination} & \multicolumn{2}{c}{spk emb} &  \multicolumn{2}{c}{model merge} \\ 
             & base speakers & interpolation & base speakers & interpolation \\
            \midrule
            Female-Female & 4.02 $\pm$ 0.14 & 4.25 $\pm$ 0.12 & 3.92 $\pm$ 0.13 & 4.24 $\pm$ 0.12\\
            Male-Male & 3.75 $\pm$ 0.13 & 3.92 $\pm$ 0.13 & 3.75 $\pm$ 0.14 & 3.91 $\pm$ 0.14\\
            Male-Female & 3.99 $\pm$ 0.12 & 3.75 $\pm$ 0.14 & 4.03 $\pm$ 0.12 & 3.45 $\pm$ 0.13 \\
            \toprule
            \multicolumn{5}{c}{WER} \\
            \toprule
             \multirow{2}{*}{Speaker combination} & \multicolumn{2}{c}{spk emb} &  \multicolumn{2}{c}{model merge} \\ 
             & base speakers & interpolation & base speakers & interpolation \\
            \midrule
            Female-Female & 11.2 & 11.5 & 12.9 & 10.9 \\
            Male-Male & 10.7 & 10.4 & 9.3 & 9.5 \\
            Male-Female & 11.0 & 10.7 & 11.1 & 11.2 \\
            \bottomrule
        \end{tabular}
    }
    \label{table:mos_evaluation}
    \end{center}
          \vspace{-10pt} %
\end{table}

Table~\ref{table:mos_evaluation} shows the evaluation results of MOS and WER.
The MOS scores indicate that the output of the merged model between same-gender base models (Female-Female, Male-Male) achieved comparable naturalness with that of single-speaker base models and speaker-embedding conditioned baseline models.
The WER results show that the generated speech of the merged model for all combinations can maintain high intelligibility compared to each base model and baseline.
This suggests that the model merging method can change only the speaker's identity while keeping the linguistic content.
In contrast, the MOS of the merged model of the different-gender combination (Female-Male) was deteriorated from base models.
\mmedit{We think that the deterioration is caused by the pitch normalization in the variance adapter of CFS2.
Since the target pitch of the variance adaptor was normalized depending on each speaker's statistic value in the CFS2 implementation we used, it is not guaranteed that the intermediate output value of the pitch predictor represents the intermediate pitch height.}

\subsubsection{Speaker interpolation smoothness}
\label{sec:secs_evaluation}

\begin{figure}[t]
\captionsetup{font={footnotesize}}
  \begin{minipage}[b]{0.45\linewidth}
    \centering
    \includegraphics[width=\hsize]{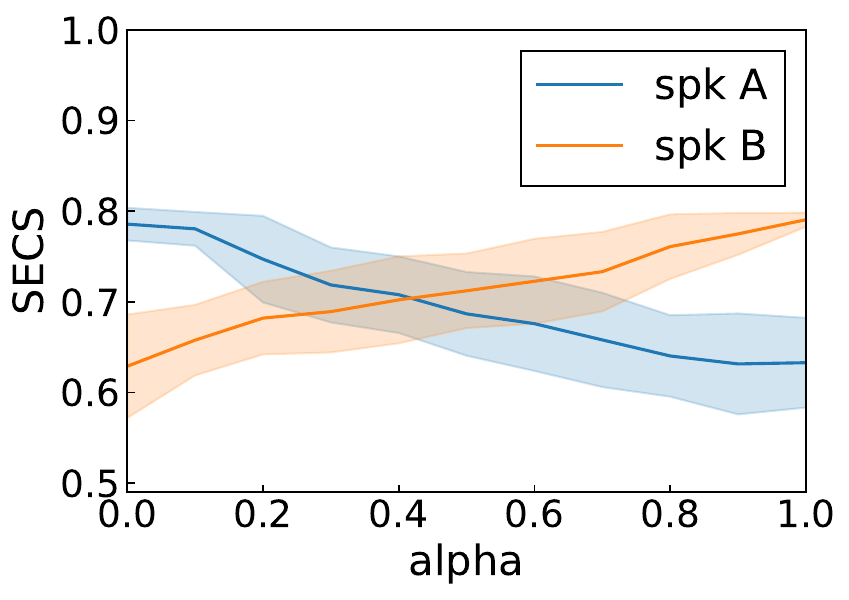}
    \subcaption{Female-Female (spk emb)}
  \end{minipage}
  \begin{minipage}[b]{0.45\linewidth}
    \centering
    \includegraphics[width=\hsize]{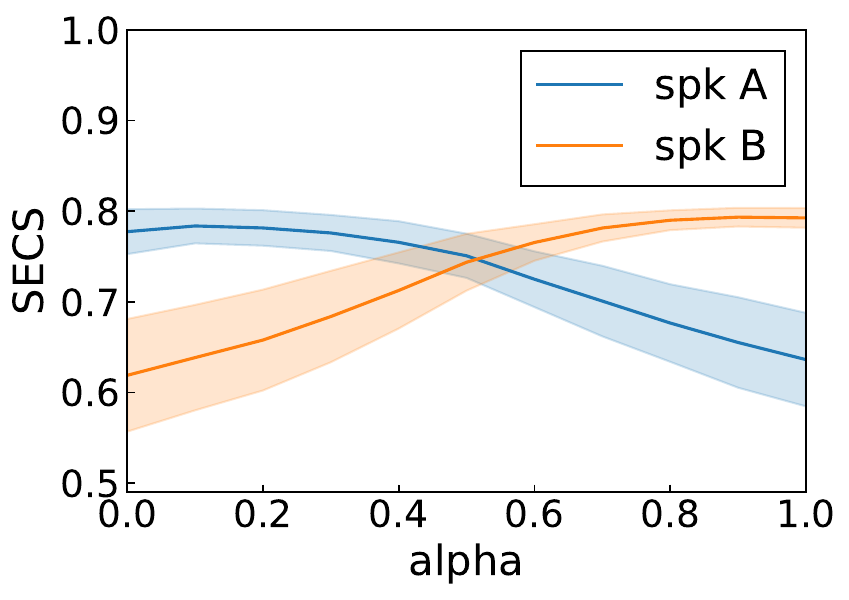}
    \subcaption{Female-Female (model merge)}
  \end{minipage}
  \begin{minipage}[b]{0.45\linewidth}
    \centering
    \includegraphics[width=\hsize]{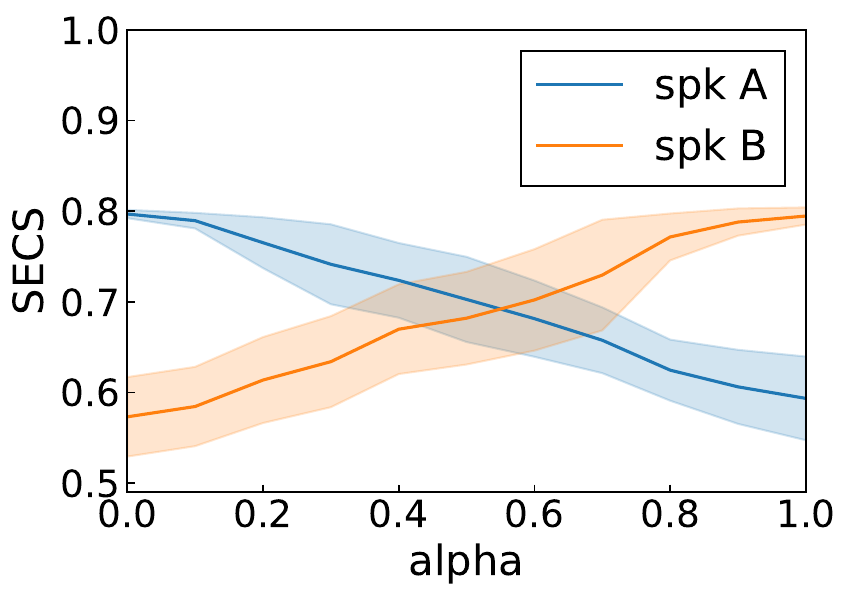}
    \subcaption{Male-Male (spk emb)}
    \end{minipage}
  \begin{minipage}[b]{0.45\linewidth}
    \centering
    \includegraphics[width=\hsize]{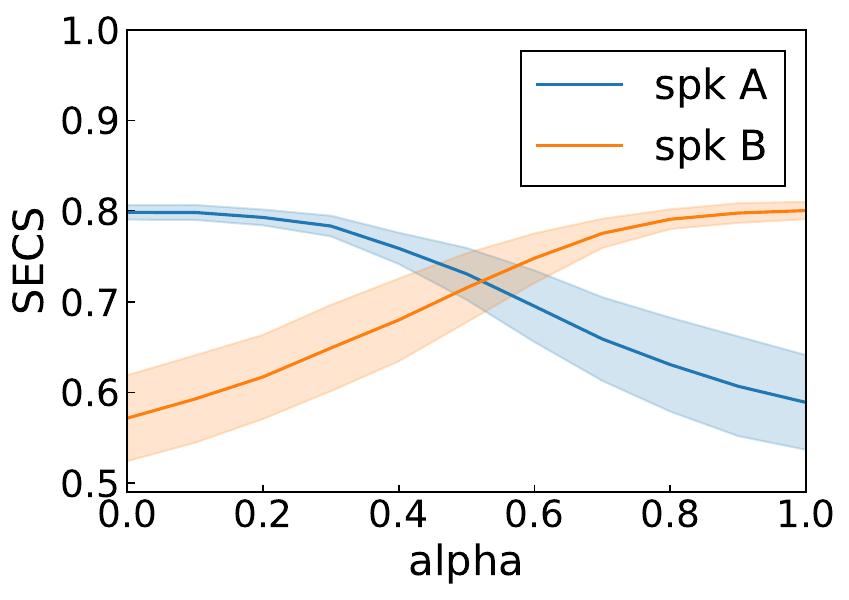}
    \subcaption{Male-Male (model merge)}
  \end{minipage}
  \begin{minipage}[b]{0.45\linewidth}
    \centering
    \includegraphics[width=\hsize]{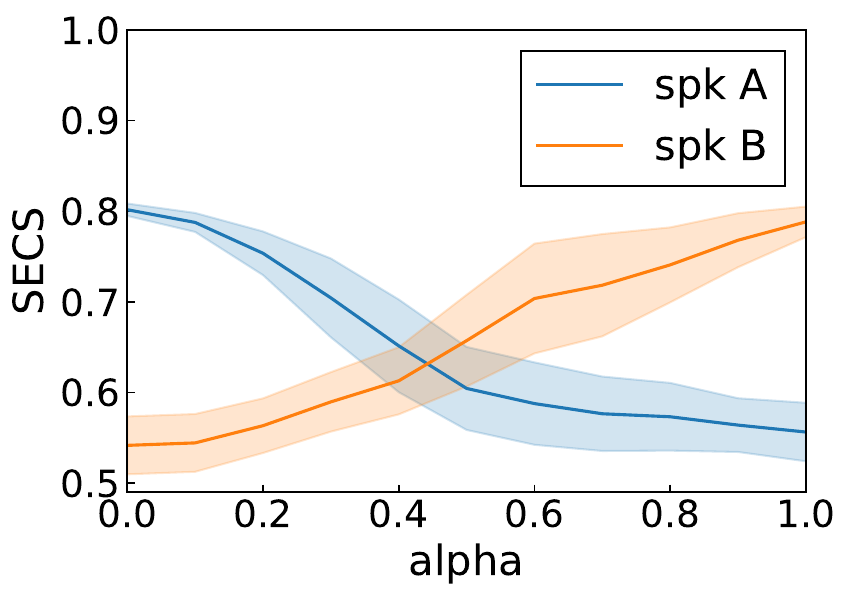}
    \subcaption{Male-Female (spk emb)}
  \end{minipage}
  \begin{minipage}[b]{0.45\linewidth}
    \centering
    \includegraphics[width=\hsize]{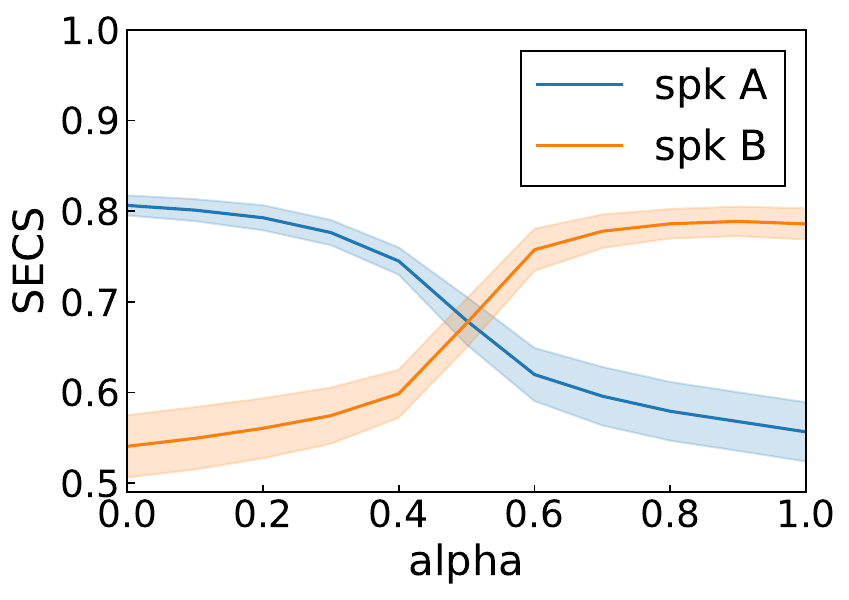}
    \subcaption{Male-Female (model merge)}
  \end{minipage}
  \centering
  \caption{SECS score with 95\% confidence
interval (CI) between the x-vectors of synthesized speech of the interpolated speaker and base speakers.
Each row corresponds to the SECS in Female-Female, Male-Male, and Male-Female gender combination settings.
The left column is the results of speaker embedding conditioned baseline model and the right column is that of model merging.}
  \label{fig:vctk_secs_score}
\end{figure}

To evaluate the smoothness of speaker interpolation by model merging, we evaluated the speaker encoder cosine similarity (SECS) score between the average x-vector of the synthetic speeches generated by the merged model and that of the ground truth (GT) speech of the original base speaker.
We merged each base model while changing the merging coefficient $\alpha$ in range of 0 to 1 in steps of 0.1 and synthesized speech by using the merged models with 10 sentences from the VCTK test set.
Figure~\ref{fig:vctk_secs_score} shows the SECS values on each gender base model combination: Female-Female, Male-Male, and Female-Male.
For all combinations, we observed that the SECS scores of the proposed method changed as smoothly as those of the baseline.
The results indicate that the model merging method can easily generate various nonexistent speakers with intermediate characteristics of any two base speakers.

\subsection{Emotion intensity control}
\label{sec:emotion_interpolation}

To investigate the controllability of the proposed method, we conducted experiments on the emotion intensity control task. 
We merged each emotion-style TTS model with a Neutral style model to control the intensity of emotion in synthesized speech. 
In this experiment, we first created each base model by using speech data from \mmedit{the five speakers 
with the five emotions described in Section~\ref{section:datatset}.}
The base models were obtained by fine-tuning from the same pre-trained model trained on the ESD dataset.
We merged one of the four emotion-style TTS models (except for Neutral style) with the Neutral style model of the same speaker, respectively. 
As a result, we obtained 20 types of merged models (5 speakers $\times$ 4 emotions).
To control the emotion intensity of synthesized speech, we changed the merging coefficient $\alpha$ from 0 to 1 in steps of 0.25.
We conducted a rearrangement ranking subjective evaluation.
We evaluated the controllability of the emotion intensity control task on the average of the ranks
metrics. 
In this subjective evaluation, participants listened to randomly arranged voice samples with five levels of emotion intensity and rearranged them in order of emotion intensity.
In the experiment, we asked 50 participants to rearrange 8 sample sets (each sample set contains 5 voice samples). 
Participants for this experiment were also recruited through AMT.

Table~\ref{table:average_rank_evaluation} shows the results of the average of the ranks on each emotion \mmedit{style} setting.
The results show that our proposed method achieved smooth emotion intensity control. 
However, in the Sad style \mmedit{setting, the range of evaluated average ranks was smaller} 
than that of other style settings.
This indicates that the controllability in the Sad \mmedit{style} setting is limited compared with other emotion styles settings.
We think that this is because of the diversity between the emotion style voice and the Neutral voice in the original dataset.
If the difference is small in the original dataset, it can be more difficult to control the emotion intensity. 
This suggests that the diversity of emotional expression in the training dataset is important for emotion intensity control by using our method.

\begin{table}[t!]
    \begin{center}
    \caption{The average ranking with each emotion intensity in 4 emotions settings}
    \scalebox{0.71}{
        \begin{tabular}{ cccccc } 
            \toprule
             & 1st & 2nd &  3rd & 4th & 5th \\ 
              \midrule
            Angry & 1.83 $\pm$ 0.22 & 2.08 $\pm$ 0.20 & 3.06 $\pm$ 0.17 & 3.63 $\pm$ 0.17 & 4.40 $\pm$ 0.24 \\
            Happy& 2.14 $\pm$ 0.25 & 2.33 $\pm$ 0.23 & 2.91 $\pm$ 0.21 & 3.52 $\pm$ 0.23 & 4.10 $\pm$ 0.26 \\
            Sad& 2.17 $\pm$ 0.19 & 2.69 $\pm$ 0.17 & 2.95 $\pm$ 0.13 & 3.35 $\pm$ 0.15 & 3.83 $\pm$ 0.17 \\
            Surprise& 1.56 $\pm$ 0.19 & 2.09 $\pm$ 0.17 & 2.76 $\pm$ 0.13 & 3.93 $\pm$ 0.15 & 4.65 $\pm$ 0.17 \\
            \bottomrule
        \end{tabular}
    }
    \label{table:average_rank_evaluation}
    \end{center}
      \vspace{-12pt} %
\end{table}

\section{Conclusions}
\label{sec:conclusion}

In this paper, we proposed an attribute interpolation method that merges two existing trained TTS models as base models while adjusting the merging coefficient.
We demonstrated that our proposed method achieves \mmedit{smooth attribute interpolation 
while maintaining the linguistic content.}

\mmedit{We conducted subjective and objective experiments in speaker generation and emotion intensity control tasks to evaluate the attribute interpolation performance of the proposed method.}
In the speaker generation experiment, we merged two single-speaker TTS models as base models with the same/different gender combination settings.
\mmedit{As a result, though our proposed method does not require specific modules for attribute interpolation}, the merged model between same-gender base models achieved comparable performance to base models and the baseline model \mmedit{using an external module.}
In the emotion intensity control experiment, we verified the controllability of the emotion intensity in synthesized speech by merging each emotion style model and the Neutral emotion style model.
As a result, we could control the intensity of each emotion smoothly by using our proposed method.

\mmedit{As a future work,} high-quality speaker interpolation between different gender settings and a multi-attribute interpolation method, such as speaker and emotion, are required for application usage such as character voice creation. 
Therefore, we plan to \mmedit{develop a more robust attribute interpolation method 
and 
a multi-attribute interpolation methods.}

\vfill\pagebreak

\bibliographystyle{IEEEtran}
\bibliography{mybib}

\end{document}